\documentclass{IEEEtran}
\ifCLASSINFOpdf
\else
\fi
\usepackage[cmex10]{amsmath}

\usepackage{amssymb}
\usepackage{pifont}
\newcommand{\cmark}{\ding{51}}%
\newcommand{\xmark}{\ding{55}}%
\usepackage{tikz}
\usepackage{cite}
\usepackage{flushend}
\usepackage{arydshln}
\usepackage{multirow}

\begin{document}
\newcommand{\convs}{cv}
\newcommand{\acs}{ac}
\newcommand{\dcs}{dc}
\newcommand{\transfs}{tf}
\newcommand{\reactor}{pr}
\newcommand{\filter}{f}

\newcommand{\Pg}{P_{g}}
\newcommand{\Qg}{Q_{g}}
\newcommand{\Pgmax}{P_{max}}
\newcommand{\Pgmin}{P_{min}}
\newcommand{\Qgmax}{Q_{max}}
\newcommand{\Qgmin}{Q_{min}}


\newcommand{\TopoDC}{\mathcal{T}^{\text{\dcs}}}
\newcommand{\TopoDCrev}{\mathcal{T}^{\text{\dcs,rev}}}
\newcommand{\TopoAC}{\mathcal{T}^{\text{\acs}}}
\newcommand{\Topoconv}{\mathcal{T}^{\text{\convs}}}
\newcommand{\Topoconvrev}{\mathcal{T}^{\text{\convs,rev}}}
\newcommand{\TopogenAC}{\mathcal{T}^{\text{gen,\acs}}}
\newcommand{\TopoloadAC}{\mathcal{T}^{\text{load,\acs}}}
\newcommand{\TopoloadDC}{\mathcal{T}^{\text{load,\dcs}}}

\newcommand{\Plij}{P^{\text{ac}}_{lij}}
\newcommand{\Plji}{P^{\text{ac}}_{lji}}
\newcommand{\Qlij}{Q^{\text{ac}}_{lij}}
\newcommand{\Qlji}{Q^{\text{ac}}_{lji}}

\newcommand{\Ui}{U_{i}}
\newcommand{\Umagi}{U^{\text{mag}}_{i}}
\newcommand{\thetai}{\theta_{i}}
\newcommand{\Uj}{U_{j}}
\newcommand{\Umagj}{U^{\text{mag}}_{j}}
\newcommand{\thetaj}{\theta_{j}}
\newcommand{\Uimax}{U^{\text{max}}_{i}}
\newcommand{\Uimin}{U^{\text{min}}_{i}}

\newcommand{\Wii}{W_{i}}
\newcommand{\Wjj}{W_{j}}
\newcommand{\Rtfij}{R^{\text{\transfs}}_{ic}}
\newcommand{\Ttfij}{T^{\text{\transfs}}_{ic}}
\newcommand{\itfc}{i^{\text{sq,\transfs}}_{c}}
\newcommand{\iprc}{i^{\text{sq,\reactor}}_{c}}

\newcommand{\Wfiltmagi}{W^{\text{\filter}}_{c}}
\newcommand{\Wconvmagi}{W^{\text{\convs}}_{c}}
\newcommand{\Uconvmagimax}{U^{\text{\convs,max}}_{c}}
\newcommand{\Uconvmagimin}{U^{\text{\convs,min}}_{c}}

\newcommand{\Rprij}{R^{\text{\reactor}}_{c}}
\newcommand{\Tprij}{T^{\text{\reactor}}_{c}}


\newcommand{\Pdcdef}{P^{\text{\dcs}}_{d^{\phi}e^{\phi}f^{\phi}}}
\newcommand{\Pdcdefo}{P^{\text{\dcs}}_{d^{0}e^{0}f^{0}}}
\newcommand{\Pdcdfe}{P^{\text{\dcs}}_{d^{\phi}f^{\phi}e^{\phi}}}
\newcommand{\Pdclossd}{P^{\text{\dcs,loss}}_{d^{\phi}}}
\newcommand{\Pdcratedd}{P^{\text{\dcs,rated}}_{d^{\phi}}}
\newcommand{\Idcdef}{I^{\text{\dcs}}_{d^{\phi}e^{\phi}f^{\phi}}}
\newcommand{\Idcdfe}{I^{\text{\dcs}}_{d^{\phi}f^{\phi}e^{\phi}}}
\newcommand{\Idcdrated}{I^{\text{\dcs,rated}}_{d^{\phi}}}
\newcommand{\Idcdmax}{I^{\text{\dcs,max}}_{d^{\phi}}}
\newcommand{\Idcdmin}{I^{\text{\dcs,min}}_{d^{\phi}}}
\newcommand{\idcdef}{i^{\text{sq,\dcs}}_{d^{\phi}e^{\phi}f^{\phi}}}
\newcommand{\Ue}{U^{\text{\dcs}}_{e^{\phi}}}
\newcommand{\Uemin}{U^{\text{\dcs,max}}_{e^{\phi}}}
\newcommand{\Uemax}{U^{\text{\dcs,max}}_{e^{\phi}}}
\newcommand{\Uf}{U^{\text{\dcs}}_{f^{\phi}}}
\newcommand{\gd}{g_{d^{\phi}}}
\newcommand{\rd}{r_{d^{\phi}}}
\newcommand{\pd}{p_{d^{\phi}}}
\newcommand{\rdo}{r_{d0}}
\newcommand{\rg}{r_{g}}
\newcommand{\Ueo}{U^{\text{\dcs}}_{e^{0}}}

%
%
\newcommand{\Pconvac}{P^{\text{\convs,\acs}}_{c^{\rho}}}
\newcommand{\Pconvacp}{P^{\text{\convs,\acs}}_{c^{1}}}
\newcommand{\Pconvacn}{P^{\text{\convs,\acs}}_{c^{2}}}
\newcommand{\Pconvacmin}{P^{\text{\convs,\acs,min}}_{c^{\rho}}}
\newcommand{\Pconvacmax}{P^{\text{\convs,\acs,max}}_{c^{\rho}}}
\newcommand{\Qconvac}{Q^{\text{\convs,\acs}}_{c^{\rho}}}
\newcommand{\Qconvacmin}{Q^{\text{\convs,\acs,min}}_{c^{\rho}}}
\newcommand{\Qconvacmax}{Q^{\text{\convs,\acs,max}}_{c^{\rho}}}
\newcommand{\Sconvacrated}{S^{\text{\convs,\acs,rated}}_{c^{\rho}}}
\newcommand{\Pconvdc}{P^{\text{\convs,\dcs}}_{c^{\rho}}}
\newcommand{\Pconvdcp}{P^{\text{\convs,\dcs}}_{c^{1}}}
\newcommand{\Pconvdcpo}{P^{\text{\convs,\dcs}}_{c^{10}}}
\newcommand{\Pconvdcn}{P^{\text{\convs,\dcs}}_{c^{2}}}
\newcommand{\Pconvdcno}{P^{\text{\convs,\dcs}}_{c^{20}}}
\newcommand{\Pconvdco}{P^{\text{\convs,\dcs}}_{c^{0}}}
\newcommand{\Pconvdcmin}{P^{\text{\convs,\dcs,min}}_{c^{\rho}}}
\newcommand{\Pconvdcmax}{P^{\text{\convs,\dcs,max}}_{c^{\rho}}}
\newcommand{\Pconvloss}{P^{\text{\convs,loss}}_{c^{\rho}}}
\newcommand{\Pconvlossp}{P^{\text{\convs,loss}}_{c^{1}}}
\newcommand{\Pconvlossn}{P^{\text{\convs,loss}}_{c^{2}}}
\newcommand{\Iconvmag}{I^{\text{\convs,mag}}_{c^{\rho}}}
\newcommand{\Iconvmaglin}{I^{\text{lin,\convs,mag}}_{c^{\rho}}}
\newcommand{\Iconvmagsq}{i^{\text{sq,\convs,mag}}_{c^{\rho}}}
\newcommand{\Iconvrated}{I^{\text{\convs,rated}}_{c^{\rho}}}
\newcommand{\aconv}{a^{\text{\convs}}_{c^{\rho}}}
\newcommand{\bconv}{b^{\text{\convs}}_{c^{\rho}}}
\newcommand{\cconv}{c^{\text{\convs}}_{c^{\rho}}}

\newcommand{\Iconvdcmag}{I^{\text{\convs,dc,mag}}_{c^{\rho}}}
\newcommand{\Iconvdcmin}{I^{\text{\convs,dc,min}}_{c^{\rho}}}
\newcommand{\Iconvdcmax}{I^{\text{\convs,dc,max}}_{c^{\rho}}}

\newcommand{\Iconvdc}{I^{\text{\convs,\dcs}}_{c^{\rho}}}
\newcommand{\Iconvdcp}{I^{\text{\convs,\dcs}}_{c^{1}}}
\newcommand{\Iconvdcpo}{I^{\text{\convs,\dcs}}_{c^{10}}}
\newcommand{\Iconvdcn}{I^{\text{\convs,\dcs}}_{c^{2}}}
\newcommand{\Iconvdcno}{I^{\text{\convs,\dcs}}_{c^{20}}}
\newcommand{\Iconvdco}{I^{\text{\convs,\dcs}}_{c^{0}}}
\newcommand{\Iconvdcg}{I^{\text{\convs,\dcs}}_{c_{g}}}

\newcommand{\Ufilti}{U^{\text{\filter}}_{c^{\rho}}}
\newcommand{\Ufiltmagi}{U^{\text{\filter,mag}}_{c^{\rho}}}
\newcommand{\thetafilti}{\theta^{\text{\filter}}_{c^{\rho}}}

\newcommand{\Uconvi}{U^{\text{\convs}}_{c^{\rho}}}
\newcommand{\Uconvmagi}{U^{\text{\convs,mag}}_{c^{\rho}}}
\newcommand{\thetaconvi}{\theta^{\text{\convs}}_{c^{\rho}}}

\newcommand{\Pl}{P_{m}}
\newcommand{\Ql}{Q_{m}}
\newcommand{\Bi}{b^{\text{shunt}}_{i}}
\newcommand{\Gi}{g^{\text{shunt}}_{i}}


\newcommand{\Ztf}{z^{\text{\transfs}}_{c^{\rho}}}
\newcommand{\Rtf}{r^{\text{\transfs}}_{c^{\rho}}}
\newcommand{\Xtf}{x^{\text{\transfs}}_{c^{\rho}}}
\newcommand{\Ytf}{y^{\text{\transfs}}_{c^{\rho}}}
\newcommand{\Gtf}{g^{\text{\transfs}}_{c^{\rho}}}
\newcommand{\Btf}{b^{\text{\transfs}}_{c^{\rho}}}

\newcommand{\Ptf}{P^{\text{\transfs}}_{c^{\rho}ie^{\rho}}}
\newcommand{\Qtf}{Q^{\text{\transfs}}_{c^{\rho}ie^{\rho}}}
\newcommand{\Ptfei}{P^{\text{\transfs}}_{c^{\rho}ie^{\rho}}}
\newcommand{\Qtfei}{Q^{\text{\transfs}}_{c^{\rho}ie^{\rho}}}
\newcommand{\Itf}{I^{\text{\transfs}}_{c^{\rho}}}
\newcommand{\Ptfp}{P^{\text{\transfs}}_{c^{1}ie^{1}}}
\newcommand{\Ptfn}{P^{\text{\transfs}}_{c^{2}ie^{2}}}
\newcommand{\Qtfp}{Q^{\text{\transfs}}_{c^{1}ie^{1}}}
\newcommand{\Qtfn}{Q^{\text{\transfs}}_{c^{2}ie^{2}}}

\newcommand{\Zpr}{z^{\text{\reactor}}_{c^{\rho}}}
\newcommand{\Rpr}{r^{\text{\reactor}}_{c^{\rho}}}
\newcommand{\Xpr}{x^{\text{\reactor}}_{c^{\rho}}}
\newcommand{\Ypr}{y^{\text{\reactor}}_{c^{\rho}}}
\newcommand{\Gpr}{g^{\text{\reactor}}_{c^{\rho}}}
\newcommand{\Bpr}{b^{\text{\reactor}}_{c^{\rho}}}
\newcommand{\Ppr}{P^{\text{\reactor}}_{c^{\rho}ie^{\rho}}}
\newcommand{\Qpr}{Q^{\text{\reactor}}_{c^{\rho}ie^{\rho}}}
\newcommand{\Pprei}{P^{\text{\reactor}}_{c^{\rho}ie^{\rho}}}
\newcommand{\Qprei}{Q^{\text{\reactor}}_{c^{\rho}ie^{\rho}}}
\newcommand{\Ipr}{I^{\text{\reactor}}_{c^{\rho}}}
\newcommand{\Bf}{b^{\text{\filter}}_{c^{\rho}}}
\newcommand{\Qf}{Q^{\text{\filter}}_{c^{\rho}}}
\newcommand{\tc}{t_{c^{\rho}}}
\include{model_components}


\title{Unbalanced OPF Modelling for Mixed Monopolar and Bipolar HVDC Grid Configurations}

\author{
Chandra~Kant~Jat,~\IEEEmembership{Member,~IEEE,}
Jay~Dave,~\IEEEmembership{Member,~IEEE,}
Dirk~Van~Hertem,~\IEEEmembership{Senior Member,~IEEE,}
Hakan~Ergun,~\IEEEmembership{Senior Member,~IEEE,}

    \thanks{This paper has received support from the NEPTUNE project, the Belgian Energy Transition fund(Corresponding author: Chandra Kant Jat.)}
    \thanks{The authors are with the Research Group ELECTA, Department of Electrical Engineering, Katholieke Universiteit, Leuven 3000 Leuven, Belgium, and also with EnergyVille 3600 Genk, Belgium (e-mail: chandrakant.jat@kuleuven.be).} 
    }
%

\maketitle

\begin{abstract}
HVDC is a critically important technology for the large-scale integration of renewable resources such as offshore wind farms. Currently, only point-to-point and multi-terminal HVDC connections exist in real-life operation. However, with the advancement of VSC-based converter technologies, future HVDC systems are foreseen to develop into  meshed HVDC grids. Bipolar HVDC grids can be operated in an unbalanced way during single pole outages or in form of mixed monopolar and bipolar grids. However, currently, there are no (optimal) power flow tools to study the feasibility of such systems. Therefore, we develop an optimal power flow (OPF) model for hybrid AC-DC grids to capture the DC side unbalances and allowing to efficiently plan and operate such future grids. In this paper, we present a multi-conductor OPF model with separate modeling of the positive pole, negative pole, metallic return conductors, and ground return. The capabilities of the model are demonstrated on a small test case, including monopolar tapping over a bipolar DC link. It is demonstrated that the developed OPF model can capture the loop flows between the different poles in unbalanced conditions, as opposed to the existing single-wire representations in the literature. Further, numerical results are presented for multiple test cases with various system sizes, starting from an 11-bus system to a 3120-bus system to demonstrate the computational tractability of the chosen model formulation.
\end{abstract}


\section{Introduction}
 \subsection{Background and literature review}
 The High Voltage Direct Current (HVDC) technology has traditionally been used for long-distance power transmission, connections of two asynchronous areas, and under-ground/sub-sea transmission links, mostly as point-to-point connections or a few multi-terminal links. Although most present-day HVCDC links are based on LCC converter technology, with recent advancements in Voltage Source Converters (VSC) and the latest developments in DC protection, meshed HVDC grids have become viable options for the future power systems\cite{flourentzou2009vsc, van2016hvdc}. Many studies show that interconnection of different countries and remote offshore wind farms using HVDC is grids is the techno-economically most viable option \cite{PIERRI2017427, innodc, bestpaths}.

 Power flow and optimal power flow models allow to investigate the feasibility and to quantify the benefits of new operational concepts. In the context of hybrid AC/DC networks, several formulations exist to solve the Optimal Power Flow (OPF) problem \cite{beerten2012generalized, wiget2012optimal, rimez2015combined, gan2014}. A sequential AC/DC power flow algorithm is discussed in \cite{beerten2012generalized} that is further developed as a Matlab-based open-source tool in \cite{beerten2015development}, whereas \cite{wiget2012optimal} and \cite{rimez2015combined} present AC/DC OPF models for VSC-based meshed DC grids. Gan et al. \cite{gan2014} present a second-order cone programming relaxation for the OPF problem in DC networks. Meyer et al. \cite{meyer2019distributed} present a distributed OPF problem for hybrid AC/DC grids, using an improved alternating direction of multipliers method (ADMM) as well as augmented Lagrangian based alternating direction inexact Newton (ALADIN) method. The authors in \cite{mevsanovic2018robust} present a robust optimization model for the AC/DC grids while including uncertainty. Hotz et. al. \cite{hotz2019hynet} describe a python-based open-source OPF framework for mixed AC/DC grids incorporating only point-to-point and radially connected multi-terminal DC networks.  Recently, a comprehensive combined AC/DC OPF model, including meshed HVDC systems, is presented in \cite{hergun_acdc_opf} which models converter stations with transformers, filters, and phase reactors in detail. Different approximations and relaxations of the non-linear optimization problem are also discussed. However,  all the above-mentioned works  model the DC grid as a balanced system using a single conductor representation. None of them model the metallic return conductor explicitly.

\begin{figure}[b!]
 	\centering
 	\includegraphics[width=\columnwidth]{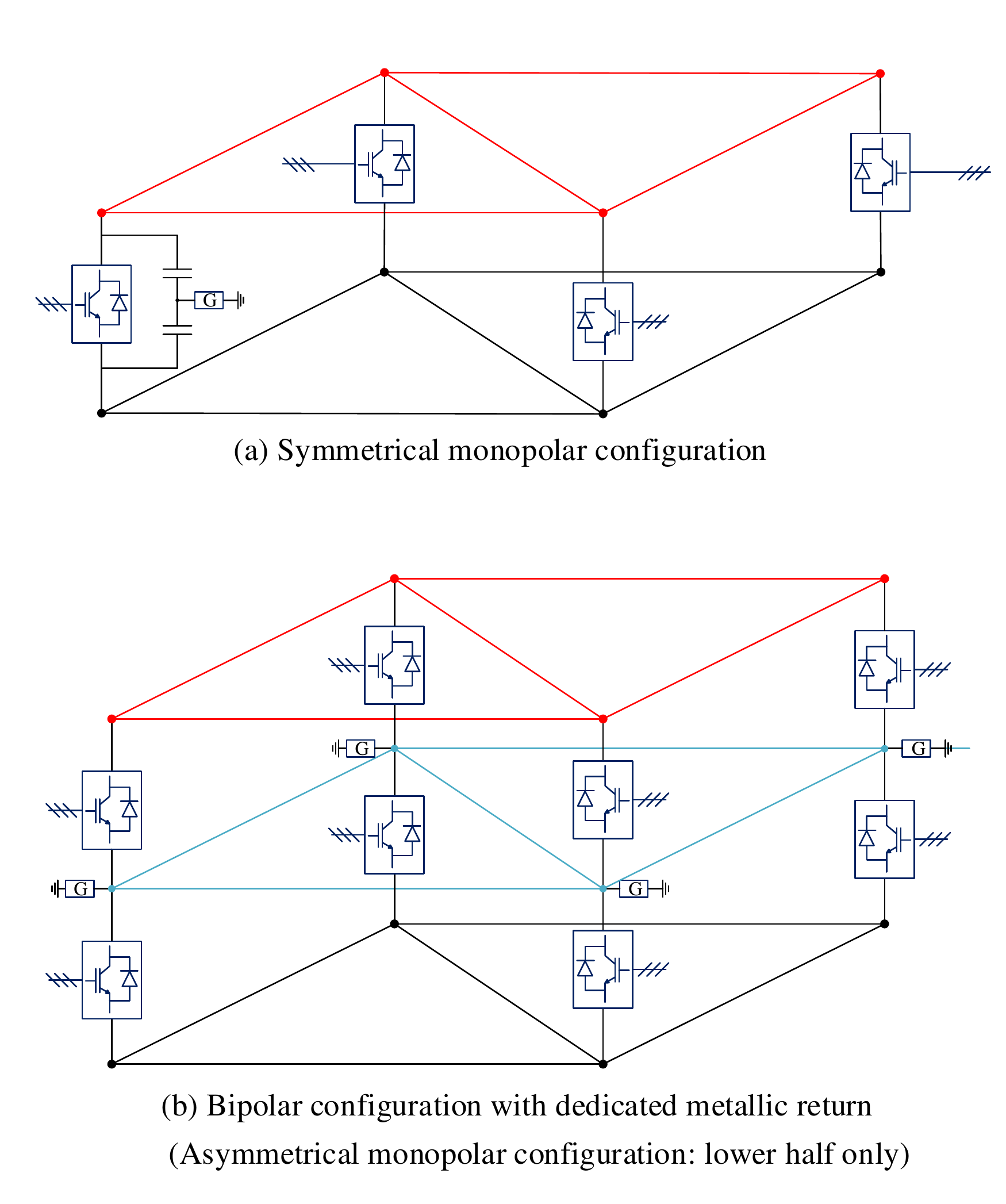}
 	\caption{HVDC grid configurations and grounding options \cite{vanhertem2019}}
 	\label{fig:HVDCConfig}
 \end{figure}

These single-conductor representations are based on the assumption that the DC grid consists of homogeneous HVDC configurations, meaning that they are carried out either as monopolar or as bipolar networks, as shown in Fig.~\ref{fig:HVDCConfig}. In the case of a monopolar DC grid, the converters and lines can have exclusively either symmetric or asymmetric configurations. Additionally, for pipolar HVDC grids, the single-conductor models assume a balanced operating state, i.e., no current flows through metallic or ground return, the same amount of power flows through the two poles of a converter, and the neutral point voltage is zero. 

 However, the existing literature on HVDC suggests that a mix of monopolar and bipolar HVDC links can help find grid designs with reduced investment costs without compromising reliability \cite{dave2019tnep, elahidoost2017expansion, dominguez2017multistage}. Moreover, with the possibility of operating the HVDC grid in unbalanced conditions, e.g. asymmetric operation of the HVDC grid, a single pole contingency at a bipolar converter station would not cause a complete terminal outage. Thus, the availability of the DC grid \cite{guo2019nodal} is improved. However, the unbalanced utilization of the positive and negative poles can cause a shift of the metallic return conductor voltage, which needs to be limited. The actual nodal voltages of the poles and metallic return depend not only on the power injections/flows in the DC grid but also on the different grounding configurations chosen at the various converter stations (Fig.~\ref{fig:HVDCConfig}b).
 
To analyze the implications of unbalanced utilization of bipolar HVDC grids,  an (optimal) power flow model is needed, which can explicitly model the DC grid unbalance. Using such an optimal power flow model, optimal positive and negative pole voltage and power set points for the HVDC converter stations can be determined for unbalanced operation while respecting AC and DC grid constraints such as current and voltage limits. In particular, it can be guaranteed that the metallic return conductor voltage does not exceed defined safety limits for various operating conditions that can occur due to the variability of renewable generation sources. 
 
 Although the multi-conductor representation of power networks is highly investigated in the context of unbalanced low voltage AC networks \cite{chen1991distribution, cheng1995three, geth2020current, geth2021real}, the unbalanced operation of HVDC networks is not investigated widely. At the same time, high-voltage AC networks are generally represented by single conductor models because the amount of unbalance among the three phases is often negligible. For asymmetric HVDC grid operation, e.g., after single pole outages of bipolar HVDC links, an unbalance of  50\%  with respect to the nominal power of the link can occur. Thus, accurately capturing the imbalance is essential for the operation of asymmetric HVDC grids. Authors in \cite{mackay2016} and \cite{mackay2018} explicitly model metallic return for bipolar grids having asymmetric loading conditions; however, with a scope limited only to  DC systems, e.g. DC-DC converter stations. Thus, to the authors' knowledge, a complete multi-conductor AC/DC OPF model for analyzing unbalanced or asymmetric HVDC grid operation does not exist in the literature.

 \subsection{Contributions and outline}
 This paper develops an optimal power flow model for unbalanced mixed monopolar/bipolar HVDC networks where both poles of bipolar converters, the transmission line, and the metallic return conductors are explicitly modeled. The grounding impedance of the HVDC converter stations, e.g. rigid grounding versus high impedance grounding, is taken into account, such that the correct voltage profile of the metallic return conductor is obtained and can be used as an optimization constraint. The AC network is modeled as a balanced network, as voltage unbalance is usually negligible in high-voltage AC networks. The HVDC converters are modeled generically, including converter station transformers, harmonic filters, and converter losses as described in \cite{hergun_acdc_opf}. 
 The power-voltage (P-V) formulation, which is the most common formulation for OPF problems, fails to capture nodal-current balance in the neutral terminal and causes numerical issues due to the low magnitude of the neutral conductor voltage. Therefore, the DC grid is modeled in terms of voltage and current variables, i.e. the I-V formulation. The developed optimization model is implemented as an extension of the open source `PowerModelsACDC.jl' package \cite{PMACDC.jl}, developed in Julia/JuMP \cite{Dun17} and uses the well-known `PowerModels.jl' package \cite{powermodels_pscc} for the modelling of the balanced AC system. The multi-conductor modeling functionality is inspired by the models included in the package `PowerModelsDistribution.jl' \cite{FOBES2020106664}. The paper's contributions with respect to the existing literature are shown in table \ref{Tab: Contribution}. 
 
The paper is outlined as follows. An OPF problem with the objective of generation cost minimization is introduced in section II. The AC/DC grid model, including the multi-conductor DC grid representation, is described in Section III. Section IV provides numerical results for a small test case and validates the developed model, whereas section V presents numerical results for larger test cases quantifying computation performance.  Finally, conclusions are presented in Section VI. 

\begin{table}[h]
\centering
\caption{Contribution}
\label{Tab: Contribution}
\begin{tabular}{lcccc}
\hline
         & AC/DC systems   & \multicolumn{2}{c}{Unbalced DC side model}     &  \\
         & including       & metallic    &  monopolar     &   open  \\
         & AC/DC  &   or ground   & {\&} bipolar  & -source \\ 
         & converter &   return   &  converter  & tool \\          \hline
\cite{beerten2012generalized, wiget2012optimal, rimez2015combined, gan2014}   &   {\cmark}     & {\xmark}      &    {\xmark}  &   {\xmark}        \\ 
\cite{beerten2015development, hergun_acdc_opf, hotz2019hynet}    &   {\cmark}     &  {\xmark}  &   {\xmark}   &  {\cmark} \\
\cite{mackay2016, mackay2018}    &   {\xmark}     &  {\cmark}  &   {\cmark}   &  {\xmark} \\
This paper &   {\cmark}     &  {\cmark}   &    {\cmark}      &  {\cmark}   \\ \hline
\end{tabular}
\end{table}
%
%
%
%
%
%
%
%
%
%
%
%
%
%
%
%

\section{Problem formulation}
\label{sec:nlmodel}

The power flow model presented in this paper is generally valid for different power system optimization objectives such as generation cost minimization, transmission loss minimization, etc. In this work we use a generation cost minimization objective as shown in \eqref{eq:objective} and the general constraints are shown throughout \eqref{eq:Ulim_ac} - \eqref{eq:conv_acdc}, which are explained in detail in the following paragraphs.
\begin{equation}
    min~\sum_{g \in \mathcal{G}} a_{g} + b_{g}\Pg + c_{g}\Pg^{2} \label{eq:objective}, 
\end{equation}
subject to
\begin{gather}
   \Uimin \leq \Ui \leq \Uimax~~~~\forall i \in \mathcal{I} \label{eq:Ulim_ac}\\
    \Uemin \leq \Ue \leq \Uemax~~~~\forall e \in \mathcal{E} \label{eq:Ulim_dc}\\
    \theta_{r} =0 ~~~~\forall r \in \mathcal{R} \label{eq:ref}\\
   \Pgmin \leq \Pg \leq \Pgmax \forall g \in \mathcal{G} \label{eq:PGlim}\\
   \Qgmin \leq \Qg \leq \Qgmin \forall g \in \mathcal{G} \label{eq:QGlim}\\
    \textnormal{AC branch flow constraint} \label{eq:branch_flow_lim_ac}\\
    \textnormal{DC branch flow constraint} \label{eq:branch_flow_lim_dc}\\
    \textnormal{AC bus KCL constraint} \label{eq:kcl_ac}\\
    \textnormal{DC bus KCL constraint} \label{eq:kcl_dc}\\
    \textnormal{AC/DC conversion constraint} \label{eq:conv_acdc}
\end{gather}


        
The objective function (\ref{eq:objective}), e.g., the active power generation costs are expressed as a quadratic cost function, where $a_{g}$, $b_{g}$ and $c_{g}$ are the corresponding cost coefficients. Equations (\ref{eq:Ulim_ac}) and (\ref{eq:Ulim_dc}) represent the AC and the DC side voltage limits, respectively, while (\ref{eq:ref}) fixes the nodal voltage angles for the reference bus(es). Active and reactive power limits of the generators are enforced in (\ref{eq:PGlim}) and (\ref{eq:QGlim}). Equations (\ref{eq:branch_flow_lim_ac}) to (\ref{eq:conv_acdc}) enforce the line flow limits, nodal current balance in AC and DC nodes, and the power flow constraints and technical limits associated with AC/DC converter stations. 
The sets and indices used throughout the paper are introduced in table~\ref{table_sets_1} whereas table~\ref{table_sets_2} presents variables and parameters of the DC grid. The well known non linear, non convex equations for modeling the balanced AC network are incorporated using the open-source tool presented in \cite{powermodels_pscc}. 
\begin{table}[h!]
    \centering
    \caption{Definition of AC/DC OPF Entities, Indices, and Sets}
    \begin{tabular}{l l}
    \hline
    {ac nodes}& $i, j \in \mathcal{I}$ \\
    {ac branches}& $l \in \mathcal{L}$ \\
    {ac topology}& $lij \in \TopoAC \subseteq \mathcal{L} \times \mathcal{I} \times \mathcal{I}$ \\
    {dc nodes} & $e, f \in \mathcal{E}$ \\
    {dc subnodes or subbranches} & $\phi \in \{1,2,0\}$ \\
    {dc branches}& $d \in \mathcal{D}$  \\
    {dc topology}& $def \in \TopoDC \subseteq\mathcal{D} \times \mathcal{E} \times \mathcal{E}$ \\
    {AC/DC converters} &$c \in \mathcal{C}$  \\
    {converter pole} &$\rho \in \{1,2\}$  \\
    {AC/DC converter topology} &$cie \in \Topoconv \subseteq\mathcal{C} \times  \mathcal{I} \times \mathcal{E}$ \\
    {generators} &$g \in \mathcal{G}$ \\
    {reference ac buses} &$r \in \mathcal{R}$ \\
    {loads} &$m \in \mathcal{M}$ \\
    {ac generator connectivity} &$gi \in \mathcal{T}^{\text{gen,\acs}} \subseteq\mathcal{G} \times \mathcal{I}$ \\
    {ac load connectivity} &$mi \in \mathcal{T}^{\text{load,\acs}} \subseteq\mathcal{M} \times \mathcal{I}$  \\
    {dc load connectivity} &$me \in \mathcal{T}^{\text{load,\dcs}} \subseteq\mathcal{M} \times \mathcal{E}$  \\ 
    \hline 
    \end{tabular}
    \label{table_sets_1}
\end{table}

\begin{table}[t]
	\renewcommand{\arraystretch}{1.3}
	\caption{DC side Parameters and variables}
	\label{table_sets_2}
	\centering
	\begin{tabular}{l l}
		\hline
		Current flow over DC branch & {$I^{dc}_{c^{\phi}}$} \\ 
		Power loss in the DC line & {$P^{d}_{loss}$} \\ 
		Current rating in the DC line & {$I^{dc,rated}_{c^{\phi}}$} \\
		Bus voltage magnitude & {$U^{dc}_{e}$}\\
		\hline 
	\end{tabular}
\end{table}

\section{AC/DC grid model}

\begin{figure*}[tbh!]
  \centering
    \includegraphics[width=1.7\columnwidth]{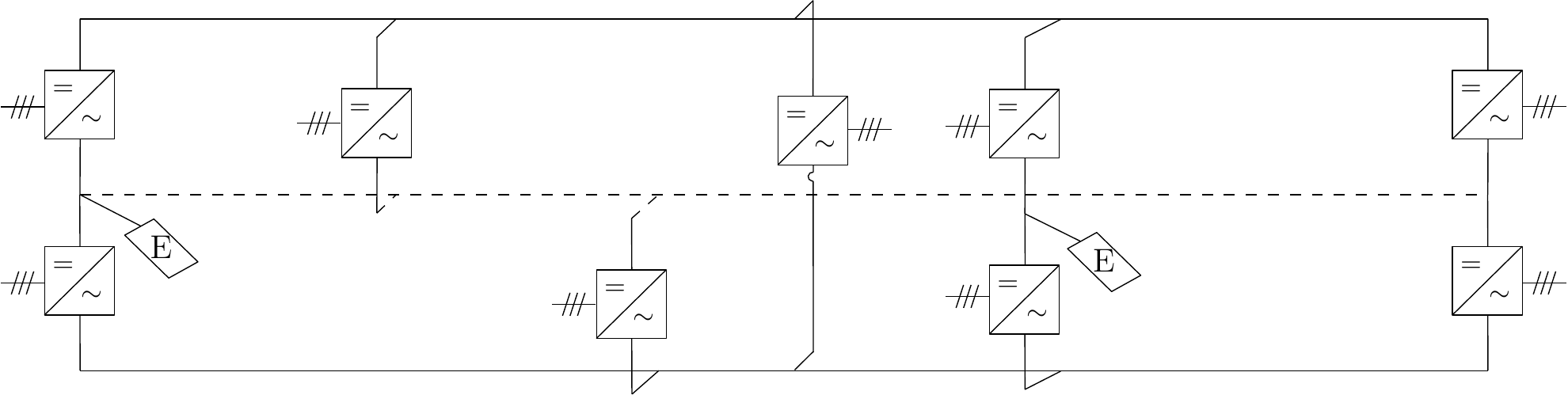} 
		\caption{An HVDC grid with a different mix of converter and line configurations (in fact, it is a multi-terminal system that can be easily extended into a meshed system). It presents a variety of tapping (two asymmetric monopoles, one symmetric monopolar, and one bipolar with ground return) on a bipolar link with metallic return \cite{leterme2014}}
		\label{fig: generic_dcgrid}
\end{figure*}

A static power flow model for the hybrid AC/DC power system is described in this section, extending the AC network equations presented in \cite{powermodels_pscc}. Although this paper's contribution lies in modeling of unbalanced DC grid, AC side modeling is also presented to show the impact of multi-conductor modeling of AC/DC converter on the AC side and to present a complete picture of the system under consideration.

\subsection{Static AC grid model}
The AC grid is modeled as a balanced system with power flow equations in the power-voltage formulation, as described in \cite{molzahn2019survey}. The active and reactive power flow through line $l$ between node $i$ and $j$ is calculated as

\begin{IEEEeqnarray}{l}
 \Plij =  g\cdot({\Umagi}^2 - {\Umagi} {\Umagj}\cdot cos(\thetai-\thetaj)) \nonumber \\
\quad -b\cdot({\Umagi}\cdot{\Umagj}\cdot sin(\thetai-\thetaj)) \quad\quad \forall lij \in \TopoAC\\
 \Qlij = -b\cdot({\Umagi}^2 - {\Umagi}\cdot{\Umagj}\cdot cos(\thetai-\thetaj)) \nonumber \\
 \quad-g\cdot({\Umagi}\cdot{\Umagj}\cdot sin(\thetai-\thetaj)) \quad\quad \forall lij \in \TopoAC
 \end{IEEEeqnarray}
Where the line admittance is given represented as $y_{ij}= g + j\cdot b$. For more details, the readers are referred to the package at \cite{powermodels_pscc}, which is directly invoked in the proposed model.  
    The nodal active and reactive balance equations on AC grid nodes $i \in \TopoAC$ are defined as: 
\begin{IEEEeqnarray}{l}
	\sum_{cie \in \Topoconv, \rho \in \{1,2\}} \Ptf + \quad\quad \!\!\!\sum_{lij \in \TopoAC}  \Plij  \nonumber \\
	= \!\! \sum_{gi \in\TopogenAC} \!\!\Pg -\!\!\!\sum_{mi \in \TopoloadAC } \!\!\Pl - \Gi (\Umagi)^2~~~\forall i \in \mathcal{I} \label{eq:kcl_acp}, \\
	\sum_{cie \in \Topoconv, \rho \in \{1,2\}} \Qtf + \sum_{lij \in \TopoAC}  \Qlij \nonumber \\
	=\!\! \sum_{gi \in \TopogenAC} \!\!\Qg -\!\!\!\sum_{mi \in  \TopoloadAC } \!\!\Ql +\Bi (\Umagi)^2~~~\forall i \in \mathcal{I} \label{eq:kcl_acq}
\end{IEEEeqnarray}.

Where $\Pg$, $\Pl$, $\Qg$, and $\Ql$ represent the active and reactive power set points of generators and loads, respectively, similarly, $\Gi+j\Bi$ is the shunt admittance connected to the AC bus.

\subsection{Static DC grid model}
A generic HVDC grid can have a mix of different configurations as shown in Fig. \ref{fig: generic_dcgrid}, originally presented in \cite{leterme2014}. It can be seen that all the different configurations consist of either one or two converter poles and either two or three conductors for DC branches. All other configurations can be represented as a subset of the bipolar converter and DC branch configurations. Therefore, the following equations are described in the context of a bipolar converter station and a bipolar HVDC link with a metallic return. \\
Since DC the grid is modeled in the I-V formulation, Kirchhoff's current law at each DC node is modeled using current variables ($\forall \phi \in {\{1,2,0\}}$):
\begin{IEEEeqnarray}{C}
\sum_{cie \in \Topoconv} \Iconvdc \!+ \sum_{cie \in \Topoconv} \Iconvdcg \!+ \sum_{def \in \TopoDC} \Idcdef = 0 ~~~~\forall e \in \mathcal{E} \label{eq:kcl_dc_c} 
\IEEEeqnarraynumspace
\end{IEEEeqnarray}
where $\Iconvdc$ is the DC converter current in pole $\rho$ connected to DC node $e$, and $\Iconvdcg$ is the converter ground current modeled as a shunt current at the neural terminal.
$\Idcdef$ is the current of DC branch $d$ connecting the DC nodes $e$ and $f$. 

\begin{figure}[tbh!]
  \centering
    \includegraphics[width=0.6\columnwidth]{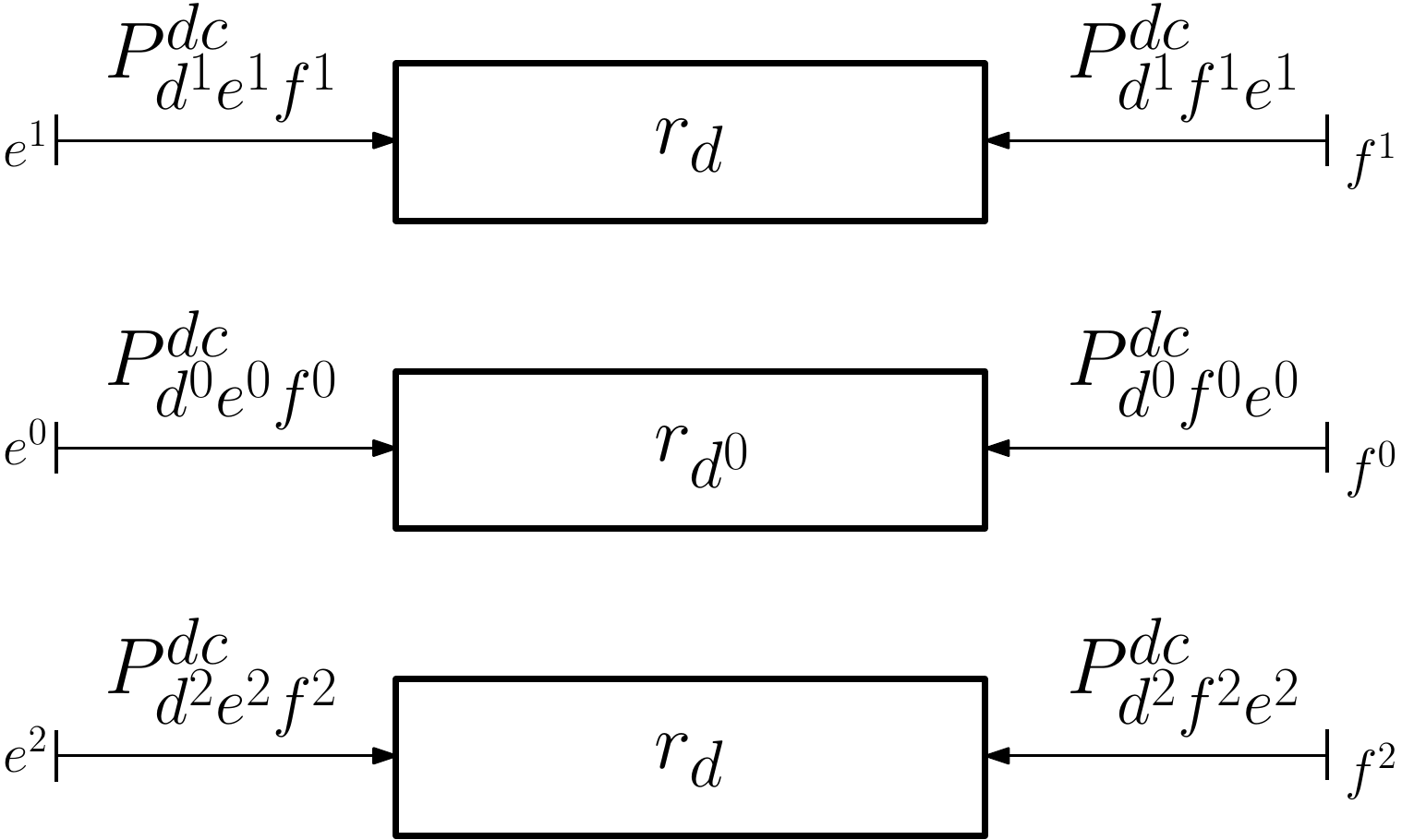} 
		\caption{Multi-conductor static model for bipolar DC branch }
		\label{fig:dcbranch}
\end{figure}

All conductors of the DC branch (i.e., positive, negative and return conductor) are modeled separately as shown in Fig. \ref{fig:dcbranch}, where $r_{d}$ is the resistance of positive pole and negative pole conductors and $r_{d0}$ is the resistance of return conductor. 

The DC grid is modeled using current and voltage variables, i.e., the I-V formulation. This choice over the commonly used P-V (power-voltage) formulation for the AC grid ensures that Kirchhoff's current law (KCL) at the neutral terminal is met. For near zero values (below numerical accuracy, e.g. tolerance of the optimization solver) of the voltage $V$  at a node, the value for power $P=V \cdot I$ also takes near zero values, thus violating the current balance in the node.     

The DC branch power flow model in terms of the current variable is defined as  
\begin{IEEEeqnarray}{C}
\Idcdef + \Idcdfe = 0 ~~~~\forall def \in \TopoDC. \label{eq:dc_If}
\end{IEEEeqnarray}

where $\Idcdef$ is the current flow in conductor $\phi$ of line $d$ in direction from node $e$ to $f$. $\Idcdfe$ is the current in the opposite direction.
The current flow for the multi-conductor line can be defined as: 
\begin{IEEEeqnarray}{C}
\Idcdef = (1/\rd) \cdot (\Ue - \Uf )~~\forall def \in \TopoDC \cup \TopoDCrev.  \label{eq:BIM}
\end{IEEEeqnarray}
where $\rd$ is the resistance of conductor $\phi$ of DC line $d$. $\TopoDCrev$ is DC topology with reverse order of nodes. It is used to apply the same equation for the opposite direction of current (or power) flow.

The branch current is constrained by the branch current limits:
\begin{IEEEeqnarray}{C}
-\Idcdrated \leq \Idcdef  \leq \Idcdrated~~\forall def \in \TopoDC \cup \TopoDCrev  \label{eq:line_limits_dc_I}. 
\end{IEEEeqnarray}

\subsection{Static AC/DC converter station model}
\begin{figure}[t!]
  \centering
    \includegraphics[width=1.0\columnwidth]{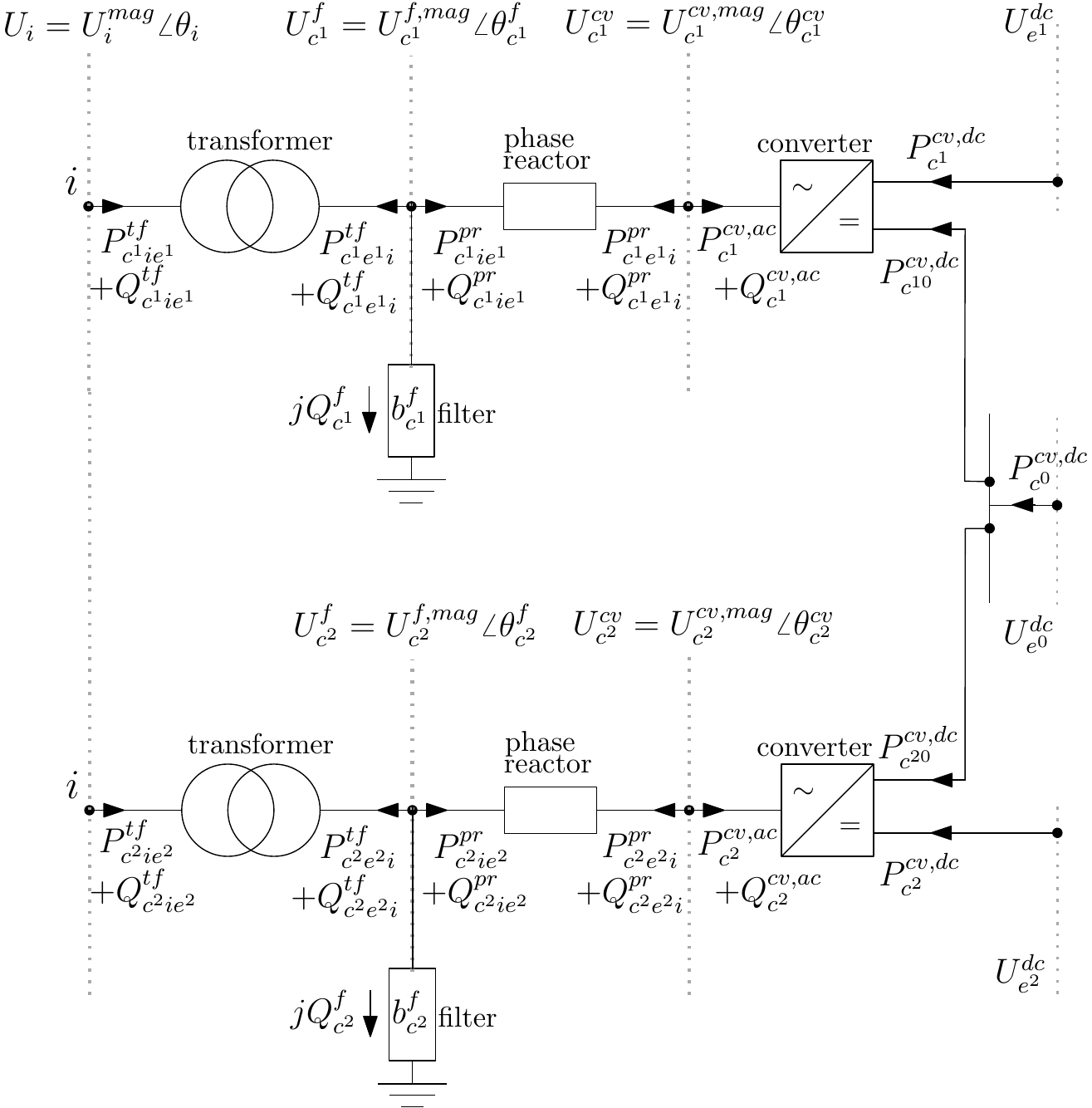} 
		\caption{Overview of the converter station model}
				\label{fig:converter}
\end{figure}

The generic AC/DC converter station model is composed of a power-electronic AC/DC converter, a phase reactor as a series impedance, a filter as a shunt susceptance, and a transformer represented with a tap ratio $t_{c}$ and a series impedance. The readers are referred to \cite{hergun_acdc_opf} for the details of the transformer, phase reactor, filter, and power electronic converter equations. These equations have been modified in this paper for the multi-conductor model.\\
 Fig.~\ref{fig:converter} depicts the bipolar converter station model and associated optimization variables. As shown in Fig.~\ref{fig:converter}, a bipolar converter station has two separate AC connections, one for the positive pole ($\rho = 1$) and one for the negative pole ($\rho = 2$). The DC side of the converter has three connections, namely,  the positive pole ($\phi = 1$), the negative pole ($\phi = 2$), and the neutral/earth ($\phi = 0$) terminal.
The AC side voltage is complex-valued, i.e., $\Ui = \Umagi \angle \thetai$, whereas the DC side voltage is real-valued represented by the voltage magnitude $\Ue$. In addition, two internal AC voltages are present at the filter connection point, $\Ufilti = \Ufiltmagi  \angle \thetafilti$, and at the power-electronic converter  $\Uconvi = \Uconvmagi  \angle \thetaconvi$. As described in \cite{hergun_acdc_opf}, any phase shift over the converter transformer is just an offset to the two internal AC voltages w.r.t. the AC bus voltage. The following paragraphs describe the different building blocks of the converter station model.

\subsubsection{AC/DC converter}
The apparent converter power is defined through the circle constraints and bound by active, reactive and apparent power limits {($\forall \rho \in {\{1,2\}}$)}:
\begin{IEEEeqnarray}{C}
(\Pconvac)^2 + (\Qconvac)^2 \leq (\Sconvacrated)^{2}~~~~\forall  cie \in \Topoconv \label{eq:conv_limits_s}, \\
(\Pconvac)^2 + (\Qconvac)^2  \!= \ (\Uconvmagi)^2 (\Iconvmag)^2~\forall  cie \in \Topoconv \label{eq:conv_loss_ac}, \\
\Pconvacmin \leq \Pconvac \leq \Pconvacmax~~~~\forall  cie \in \Topoconv \label{eq:conv_limits_p}, \\
\Qconvacmin \leq \Qconvac \leq \Qconvacmax~~~~\forall  cie \in \Topoconv\label{eq:conv_limits_q}, \\
\Uconvmagimin \leq  \Uconvmagi \leq \Uconvmagimax~~~~\forall  cie \in \Topoconv.
\end{IEEEeqnarray}
Active power injected/absorbed by the converter into/from the DC grid is modeled by the variable $\Pconvdc$ as shown in Fig.~\ref{fig:converter}. This power is bound to operational limits as 

\begin{IEEEeqnarray}{C}
\Pconvdcmin \leq \Pconvdc \leq \Pconvdcmax~~~~\forall  cie \in \Topoconv. \label{eq:conv_limits_dc_p}
\end{IEEEeqnarray}
The AC and DC side power injections of the converter poles are linked using the converter losses:
\begin{IEEEeqnarray}{C}
\Pconvac + \Pconvdc + \Pconvdco = \Pconvloss  ~~ \forall  cie \in \Topoconv. \label{eq:conv_ac_dc1}
\end{IEEEeqnarray}
Where $\Pconvdcpo$ and $\Pconvdcno$  represent the power flow between the positive and negative poles and the neutral terminal, respectively. $\Pconvlossp$ and $\Pconvlossn$ are the losses of the positive and negative poles of the converter and are defined as,
\begin{IEEEeqnarray}{C}
\!\Pconvloss = \aconv + \bconv \cdot \Iconvmag + \cconv \cdot (\Iconvmag)^2~\forall  cie \in \Topoconv \label{eq:conv_losses}.
\IEEEeqnarraynumspace
\end{IEEEeqnarray}
In \eqref{eq:conv_losses}, $\Iconvmag$ is the AC side current magnitude of pole $\rho$ of the converter $c$. This current is constrained by the converter's current rating $\Iconvrated$ for each pole $\rho \in {\{1,2\}}$:
\begin{IEEEeqnarray}{C}
\Iconvmag \leq  \Iconvrated~~~~\forall  cie \in \Topoconv. \label{eq_Iconvmag}
\end{IEEEeqnarray}
DC side converter current is linked to the DC side power injection using \eqref{eq:conv_ac_d_dc} is bound by the current rating $\Iconvdcmax$ using \eqref{eq:conv_ac_d_dc_lim}. 
\begin{IEEEeqnarray}{C}
\Pconvdc = \Ue \cdot \Iconvdcmag ~~~~\forall  cie \in \Topoconv, \label{eq:conv_ac_d_dc} \\
\Iconvdcmin \leq  \Iconvdcmag \leq \Iconvdcmax~~~~\forall  cie \in \Topoconv \label{eq:conv_ac_d_dc_lim}.
\end{IEEEeqnarray}

\subsubsection{Transformer}
The active and reactive power seen at the AC bus $i$ of the converter transformer of pole $\rho$ of converter $c$ are defined as:  
\begin{IEEEeqnarray}{C}
	\Ptf = \Gtf \left(\frac{\Umagi}{\tc} \right)^2  -\Gtf \frac{\Umagi}{\tc} \Ufiltmagi \cos(\thetai-\thetafilti)   \nonumber \\
	-\Btf \frac{\Umagi}{\tc} \Ufiltmagi \sin(\thetai-\thetafilti)~~\forall  cie \in \Topoconv \cup \Topoconvrev, \label{tr_start_p}\\
	\Qtf =-\Btf \left(\frac{\Umagi}{\tc}\right)^2 +  \Btf \frac{\Umagi}{\tc} \Ufiltmagi \cos(\thetai- \thetafilti)    \nonumber \\
	-\Gtf \frac{\Umagi}{\tc} \Ufiltmagi \sin(\thetai-\thetafilti)~~\forall  cie \in \Topoconv \cup \Topoconvrev. \label{tr_start_q}
\end{IEEEeqnarray}
where $\Gtf$ and $\Btf$ are the conductance and susceptance of the converter transformers for the $\rho^{th}$ pole, respectively. $\tc$ is the  converter transformer tap changer setting.  

\subsubsection{Capacitive filter}
The reactive power of the filter capacitor of pole  $\rho$ of the converter $c$ is calculated as:  
\begin{IEEEeqnarray}{C}
	\Qf = - \Bf (\Ufiltmagi)^2~~~~\forall  cie \in \Topoconv. 
\end{IEEEeqnarray}

\subsubsection{Phase reactor}
The phase reactor impedance is defined as $\Zpr = \Rpr + j \Xpr$, with the equivalent admittance $\Ypr = \frac{1}{\Zpr} = \Gpr + j \Bpr$. As both the transformer and the reactor are modeled as a series impedance, for modeling the phase reactor equations (\ref{tr_start_p}) - (\ref{tr_start_q}) can be used by choosing  $\tc = 1$ and using the respective voltage variables and angles according to Fig.~\ref{fig:converter}.  

The active and reactive power balance at the node connecting the transformer, the filter capacitor, and the phase reactor is defined as:
\begin{IEEEeqnarray}{C}
	\Ppr +  \Ptfei= 0  ~~\forall  cie \in \Topoconv, \label{eq_p_filter_balance} \\
	\Qpr +  \Qtfei + \Qf= 0~~\forall  cie \in \Topoconv.  \label{eq_q_filter_balance} 
\end{IEEEeqnarray}

\subsubsection{Converter DC side current constraints}
\begin{figure}[t!]
  \centering
    \includegraphics[width=.5\columnwidth]{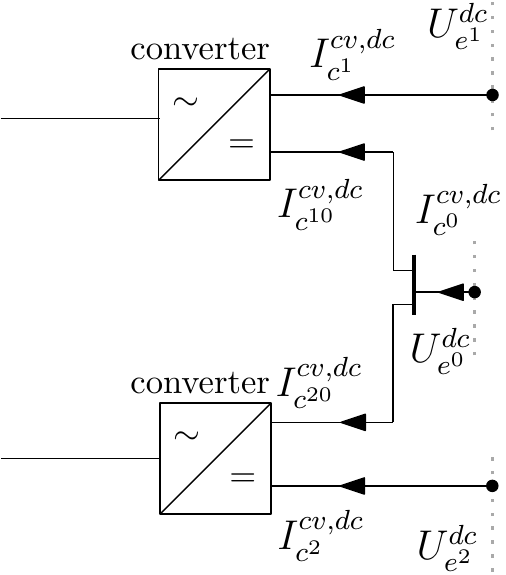} 
		\caption{Visualisation of the converter current constraints }
				\label{fig:converter_current}
\end{figure}

Fig. \ref{fig:converter_current} shows an individual pole representation of the DC side of a bipolar converter station. For converter $c$, the superscripts $1$ and $2$ indicate positive and  negative poles, respectively, whereas $0$ indicates the neutral. To ensure the numerical stability of the optimization problem, the DC side converter currents need to be modeled explicitly. Since, the power-only representation fails to model the system accurately due to the low, neutral voltage, e.g., for low-impedance grounding on the DC side. The DC side converter's current constraints are
\begin{IEEEeqnarray}{C}
\Iconvdcp + \Iconvdcpo = 0 \label{eq_Iconvdc_p}, \\
\Iconvdcn + \Iconvdcno = 0 \label{eq_Iconvdc_n},\\
\Iconvdco = \Iconvdcpo + \Iconvdcno \label{eq_Iconvdc_o},\\
\Iconvdcp+ \Iconvdcn+ \Iconvdco =0 \label{eq_Iconvdc}.
\end{IEEEeqnarray}

Equations \eqref{eq_Iconvdc_p} and \eqref{eq_Iconvdc_n} imply that the DC current entering each pole should be equal to the current exiting. Equation \eqref{eq_Iconvdc_o} defines the net current in the neutral conductor, and \eqref{eq_Iconvdc} states that the difference of positive and negative pole currents must solely flow through the neutral conductor. 

\subsubsection{DC grid grounding constraint}
DC grid is modeled in the most generic way allowing multiple grounding points. Although theoretically possible \cite{leterme2014}, the use of the ground as a return in HVDC operation is not permitted in many parts of the world \cite{vanhertem2019}. In this model, ground return operation has been included for the sake of generality. The ground is modeled as a superconducting plane, whereas the impedance of the ground return path is lumped at the grounding terminals. Thus, the ground is represented as a shunt element connected at the neutral terminal of the respective DC bus. The grounding impedance in the steady state DC model is effectively a resistance. It is indicated by the red color in Fig. \ref{fig:test_case}.  In this model, the option of grounding is considered only at the DC buses with converter stations which can be easily extended to other buses. The grounding current is defined as
\begin{IEEEeqnarray}{C}
\Iconvdcg = \frac{\Ueo}{\rg}    ~~ \forall c \in \mathcal{C} \label{eq_Iconvdcg}.
\end{IEEEeqnarray}
where $\Ueo$ is neutral terminal voltage magnitude of DC bus $e$ to which converter $c$ is connected and $\rg$ is the grounding resistance.

\section{Test case and Numerical Results}
 This section presents the impact of the multi-conductor modeling that can capture unbalanced loading conditions of  DC grids. The test case and numerical results are as follows. 
 
\subsection{Test case}
\begin{figure}[tbh]
        \centering
        \includegraphics[width=1\columnwidth]{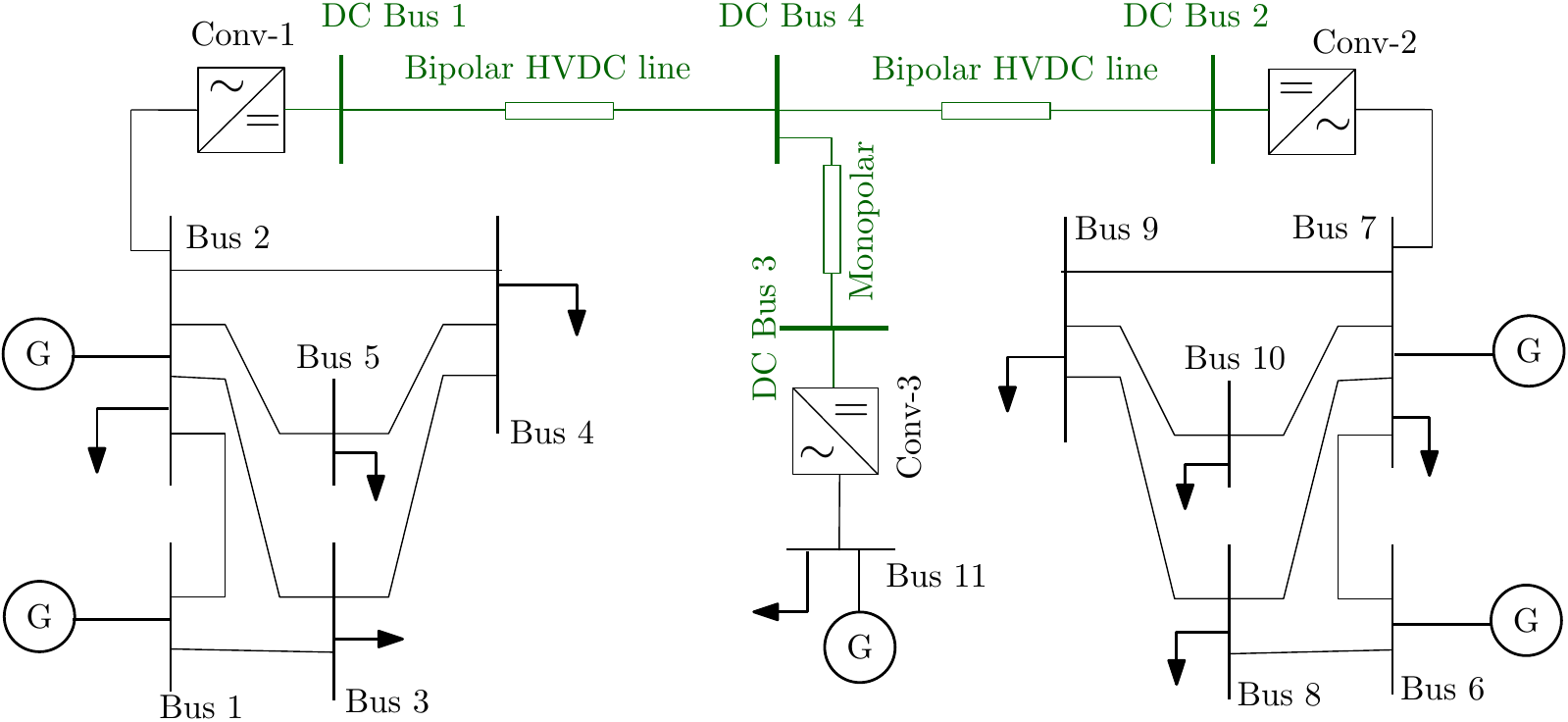}
        \caption{Single line diagram of the test case}
        \label{fig:test_case_SLD}
    \end{figure}

\begin{figure}[tbh]
     \centering
      \includegraphics[width=1\columnwidth]{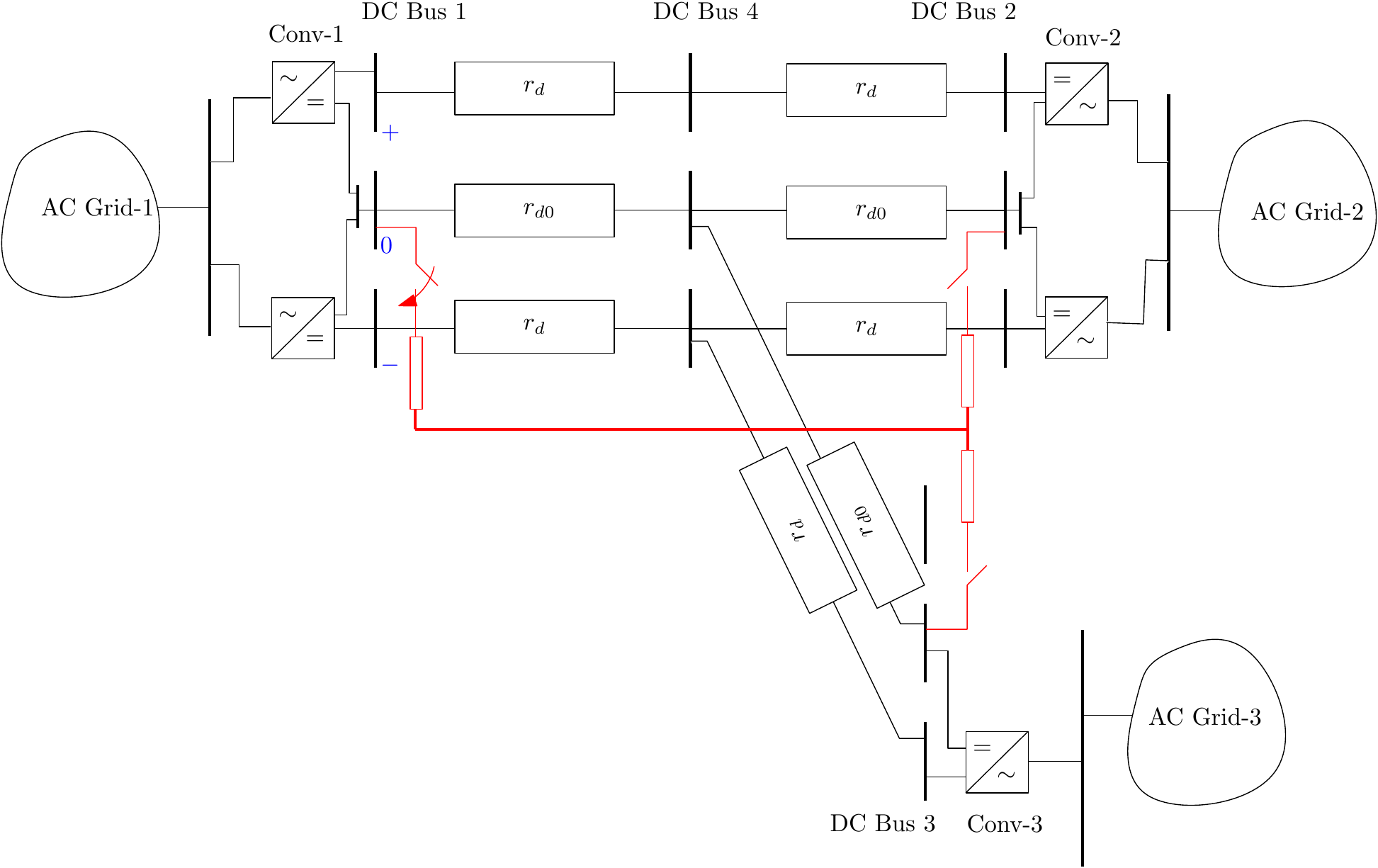} 
		\caption{Circuit diagram of the unbalanced DC network used in the test case. The red connections indicate the grounding impedances. }
    	\label{fig:test_case}
    \end{figure}

The test case consists of 3 AC systems interconnected through an HVDC system, as shown in Fig. \ref{fig:test_case_SLD}. AC grid 1 and AC grid 2 are two identical 5-bus systems, each with two generators. They are connected to the DC grid through bipolar converter stations Conv-1 and Conv-2, respectively. AC grid 3 is a single bus system with a generator and a load at the same bus. Its connection to the DC grid is through Conv-3, which has an asymmetric monopolar configuration. This monopolar converter is connected to the bipolar link through a monopolar tapping at the negative and neutral terminals of  DC bus 4. In this test case, the generation costs are modified such that AC grid 1 is the cheapest zone, whereas AC grid 3 is the costliest one. Thus, it can be expected that the OPF solution results in export from AC grid 1 and import to AC grid 3. The exact multi-conductor representation of the DC side of this test system is presented in Fig.~\ref{fig:test_case}. All results presented throughout the tables are in per unit ($pu$) for power, current, and voltages, whereas it is USD(\$) for the objective function unless specified otherwise. The power base is 100~$MVA$, whereas the voltage base is 345~$kV$.

\subsection{Validation against existing balanced formulations}
 Since there are no pre-existing benchmark models and test cases for unbalanced AC/DC networks, direct validation is not possible. A balanced network is an edge case of a generic unbalanced network, and thus, it can be used to validate the model indirectly. As the proposed multi-conductor model can equally be applied to balanced networks, its output has been compared to the balanced AC/DC OPF model of the PowerModelsACDC.jl package \cite{PMACDC.jl}.  \\
If Conv-3 and the monopolar link (between DC buses 3 and 4) in the test case (Fig.~\ref{fig:test_case}) are converted to bipolar configurations, the resulting network becomes a balanced bipolar HVDC system which can be solved using both the balanced and unbalanced multi-conductor models. By solving a generation cost minimization OPF problem using both models, it is found that the values of the objective function and generator set points match up to the solver's numerical accuracy ($1e-6$). Both AC and DC side branch flows and bus voltages are found to be the same. The converter power flows and DC grid voltage magnitudes are presented in Table \ref{table: mcdc_balanced} where $ACDC$ refers to the solution obtained with the balanced single-conductor OPF model, and  $MCDC$ refers to the results obtained with the proposed multi-conductor model. As expected, we can see that the converter powers are equally divided between both poles. Positive and negative terminal voltages are equal for each DC bus, while the neutral voltage is zero. Thus, it can be concluded the proposed $MCDC$ model, with the detailed representation of the DC side, finds correct solutions. 
\begin{table}[ht]
\centering
\caption{Comparing $ACDC$ and $MCDC$ results for balanced system}
\label{table: mcdc_balanced}
\begin{tabular}{p{0.8cm}llcccl}
\hline
 &  & \multicolumn{4}{c}{DC bus voltage magnitude} \\
 &  & Bus 1 & Bus 2 & Bus 3 & \multicolumn{1}{c}{Bus 4} \\
 $ACDC$ &  & 1.1 & 1.08134 & 1.07565 & \multicolumn{1}{c}{1.08566} \\
\hdashline
\multirow{3}{*}{$MCDC$} & Positive & 1.1 & 1.08134 & 1.07565 & \multicolumn{1}{c}{1.08566} \\
 & Negative & -1.1 & -1.08134 & -1.07565 & \multicolumn{1}{c}{-1.08566} \\
 & Neutral & 0.000 & 0.000 & 0.000 & \multicolumn{1}{c}{0.000} \\
 \hline
 &  & \multicolumn{3}{c}{DC side power injections ($MW$)} &  \\
 &  & Conv-1 & Conv-2 & Conv-3 &  \\
$ACDC$ &  & 62.53 & -16.80 & -40.00 &  \\
\hdashline
\multirow{2}{*}{$MCDC$} & Positive Pole & \multicolumn{1}{r}{31.27} & \multicolumn{1}{r}{-8.40} & \multicolumn{1}{r}{-20.00} &  \\
 & Negative pole & \multicolumn{1}{r}{31.27} & \multicolumn{1}{r}{-8.40} & \multicolumn{1}{r}{-20.00} & \\
 \hline
\end{tabular}
\end{table}
    
   \subsection{Unbalanced OPF using balanced OPF tool}
\begin{figure}[h]
     \centering
      \includegraphics[width=1\columnwidth]{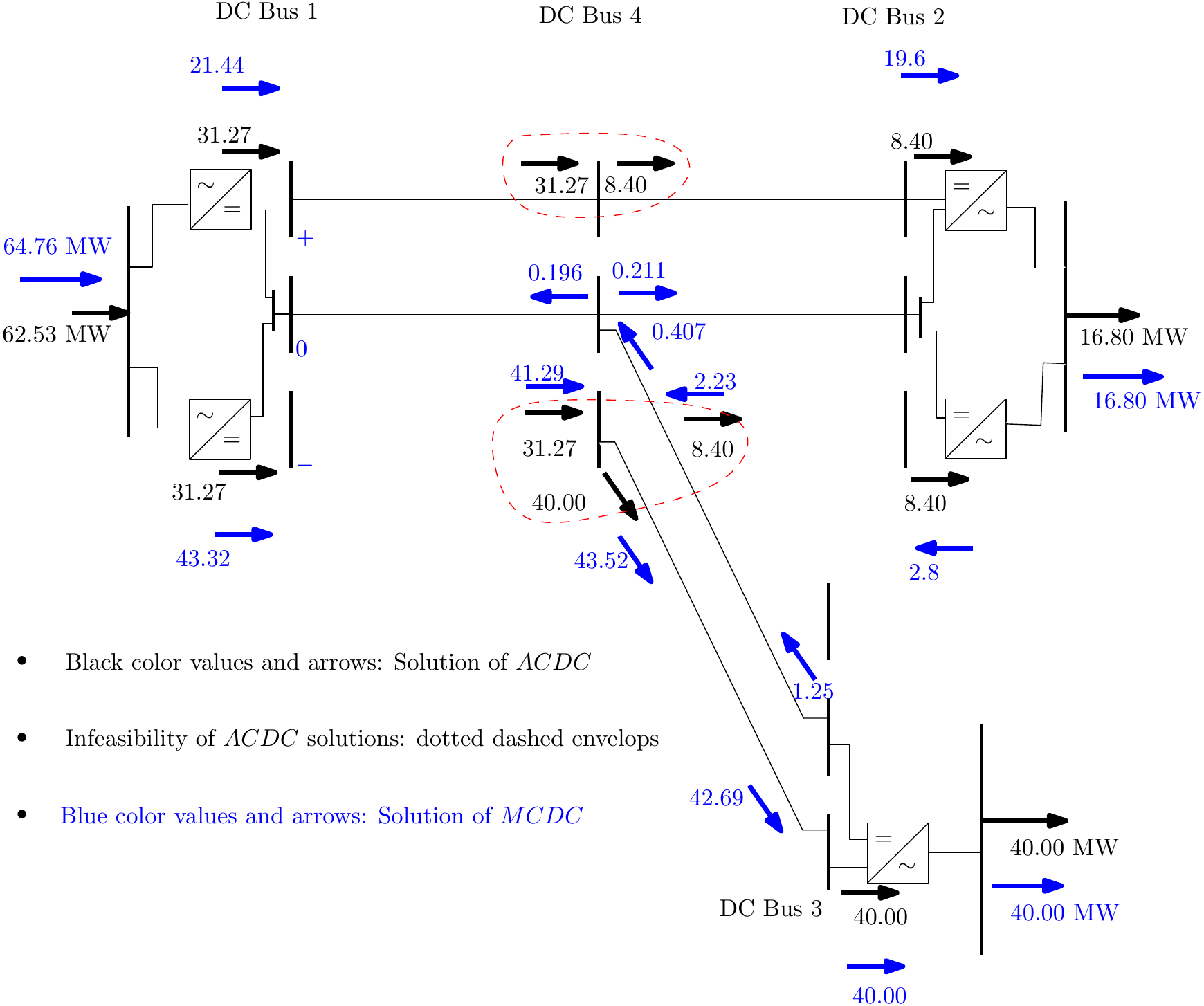}
		\caption{Infeasibility of solutions obtained with balanced OPF model ($ACDC$) and correct solutions with the proposed unbalanced OPF model ($MCDC$) }
    	\label{fig:test_case_result}
    \end{figure}

    In this section, an attempt is made to solve the original unbalanced test case with the already existing balanced OPF tool $ACDC$. The solution of the system under consideration with this balanced model results in the net power exchange as indicated in the first row of Table \ref{table: pgrid_acdc}. For the individual pole flows, such models rely on equally dividing the bipolar converter power to both poles ($2^{nd}$ and $3^{rd}$ rows of Table \ref{table: pgrid_acdc}). Although a solution is found by the optimizer because equal power division between the converter station poles is assumed, the obtained solution is infeasible in reality. This is shown in Fig. \ref{fig:test_case_result}, where the black arrows indicate the solution obtained by the balanced model. Due to the assumption of equal power sharing, the obtained solution of the balanced model does not respect the nodal power balance at positive and negative terminals of DC bus 4 (highlighted with red dashed lines). 
    Since the monopolar link is connected to the negative and neutral conductors of the  bipolar link, both poles of Conv-1 and Conv-2 would not be loaded equally. This phenomenon can not be captured by balanced AC/DC OPF models. Although the single conductor $ACDC$ model accurately captures the net power exchanges between the AC and DC grid, it fails to make an appropriate allocation of the power set points of each converter pole in operation. Thus, explicit representation of each converter pole and DC link conductor is essential.
    
    \begin{table}[!]
    \caption{Power exchange ($MW$) between AC networks and DC network using balanced optimal power flow tool}
    \centering
    \label{table: pgrid_acdc}
    \begin{tabular}{llll}
    \hline 
                              & \multicolumn{1}{l}{Conv 1} & \multicolumn{1}{l}{Conv 2} & \multicolumn{1}{l}{Conv 3} \\
        Net Power Transfer     & 62.53                          & -16.80                         & -40.00                         \\
    \hdashline
        Positive Pole  & 31.27                          & -8.40                         & -                         \\
        Negative Pole & 31.27                          & -8.40                         & -40.00                        \\
    \hline
    \end{tabular}
    \end{table}

    \subsection{Unbalanced OPF using the multi-conductor model}
    In this case, the OPF problem is solved using the proposed multi-conductor AC/DC OPF model ($MCDC$). The generator set-points obtained by the single conductor ($ACDC$) and the multi-conductor models are compared in Table \ref{table: Pg}. It can be observed that the generator set-points are the same in both cases except for a slight difference in the set-point of the generator at bus 2. This difference is caused by the non-linear losses of the converters and DC branches. As the converter and line losses depend on the square of the current, the unbalanced distribution of the currents results in higher losses, which are reflected in the higher generator output. 
    \begin{table}[!]
        \caption{Generator dispatch obtained by the balanced and multi-conductor model}
        \centering
        \label{table: Pg}
        \begin{tabular}{cccc}
        \hline
\multicolumn{1}{l}{Generator} & \multicolumn{1}{l}{AC Bus no.} & \multicolumn{1}{l}{Pg ($ACDC$)} & \multicolumn{1}{l}{Pg ($MCDC$)} \\ \hline
1                             & 1                              & 1.42883                       & 1.42883                       \\
2                             & 2                              & 0.89328                       & 0.91557                       \\
3                             & 6                              & 1.42882                       & 1.42881                       \\
4                             & 7                              & 0.10000                           & 0.10000                           \\
5                             & 11                             & 0.10000                           & 0.10000                          \\
        \hline
        \end{tabular}
    \end{table}

    However, the value of the proposed model is clear from the results provided in Tables \ref{table: dcbus_voltage} and \ref{table: mcdc_results}. Table \ref{table: dcbus_voltage} presents the p.u. voltage magnitudes of the positive, negative, and neutral terminals of the DC buses. Since the neutral point of converter 1 (DC bus 1) is grounded, its magnitude is zero. Due to the flow of the current through the neutral conductor, the neutral terminal voltage magnitudes at other buses take non-zero values. The positive and negative terminal voltages can also be observed to be in the constrained range of 0.9 to 1.1~$pu$ and are pushed towards the upper (lower in the case of the negative pole) bounds to minimize the losses in the system. 
    \begin{table}[!]
    \caption{Voltages over the terminal of DC Buses}
    \centering
    \label{table: dcbus_voltage}
    \begin{tabular}{lrrrr}
    \hline
                    & \multicolumn{1}{l}{DC Bus 1} & \multicolumn{1}{l}{DC Bus 2} & \multicolumn{1}{l}{DC Bus 3} & \multicolumn{1}{l}{DC Bus 4} \\
Postive terminal  & 1.1                          & 1.0804                       & -                            & 1.0902                       \\
Negative terminal & -1.1                         & -1.0812                      & -1.0592                      & -1.0801                      \\
Neutral terminal  & 0                            & 0.0008                       & -0.0311                      & -0.0101                     \\
    \hline
    \end{tabular}
    \end{table}

 The top section of Table \ref{table: mcdc_results} shows active power injections from the AC side to each converter pole at the corresponding converter station. It can be observed that the power flowing through the different poles differs not only in value but also the direction (sign). Here the positive pole of Conv-2 feeds power from the DC grid to the AC grid, whereas the negative pole does the opposite creating a circulating power between the positive and negative poles (see Fig. \ref{fig:test_case_result}). The difference in the power injection values at both ends of the DC branches (Fig. \ref{fig:test_case_result}) is caused by the DC branch losses. 
 The middle section of Table \ref{table: mcdc_results} shows the DC power output of the converters, which differs from the above values by the converter losses. The last section of Table \ref{table: mcdc_results} shows the current balance across the terminals of the DC converters. It can be observed that the current balance between the positive and negative poles and the neutral is conserved as intended.

\begin{table}[!]
    \caption{Values of converter variables for unbalanced OPF}
    \centering
    \label{table: mcdc_results}
    \begin{tabular}{lrrr}
    \hline
                        \multicolumn{4}{c}{Pgrid (from AC grid to conv.) $MW$}                                                                           \\
                       & \multicolumn{1}{l}{Conv 1} & \multicolumn{1}{l}{Conv 2} & \multicolumn{1}{l}{Conv 3} \\
Positive  pole  & 21.44                          & -19.6                          & -                               \\
Negative pole & 43.32                          & 2.8                           & -40.00                            \\
\hdashline
                        \multicolumn{4}{c}{Pdc (from DC grid to conv.) $MW$}                                                                             \\
                       & \multicolumn{1}{l}{Conv 1} & \multicolumn{1}{l}{Conv 2} & \multicolumn{1}{l}{Conv 3} \\
Postive terminal       & -20.7                          & 20.33                          & -                               \\
Negative terminal      & -42.05                         & -2.24                         & 42.69                          \\
Neutral terminal       & 0                               & -02.0                         & -1.25                         \\
\hdashline
                        \multicolumn{4}{c}{Idc (DC side conv. current) $pu$}                                                                             \\
                       & \multicolumn{1}{l}{Conv 1} & \multicolumn{1}{l}{Conv 2} & \multicolumn{1}{l}{Conv 3} \\
Postive terminal       & -0.1882                         & 0.1882                          & -                               \\
Negative terminal      & 0.3823                          & 0.0207                          & -0.403                          \\
Neutral terminal       & -0.1941                         & -0.2089                         & 0.403          \\
\hline
\end{tabular}
\end{table}

\subsection{Neutral point voltages with increasing system imbalance}
\begin{figure}[h]
        \centering
        \includegraphics[width=1\columnwidth]{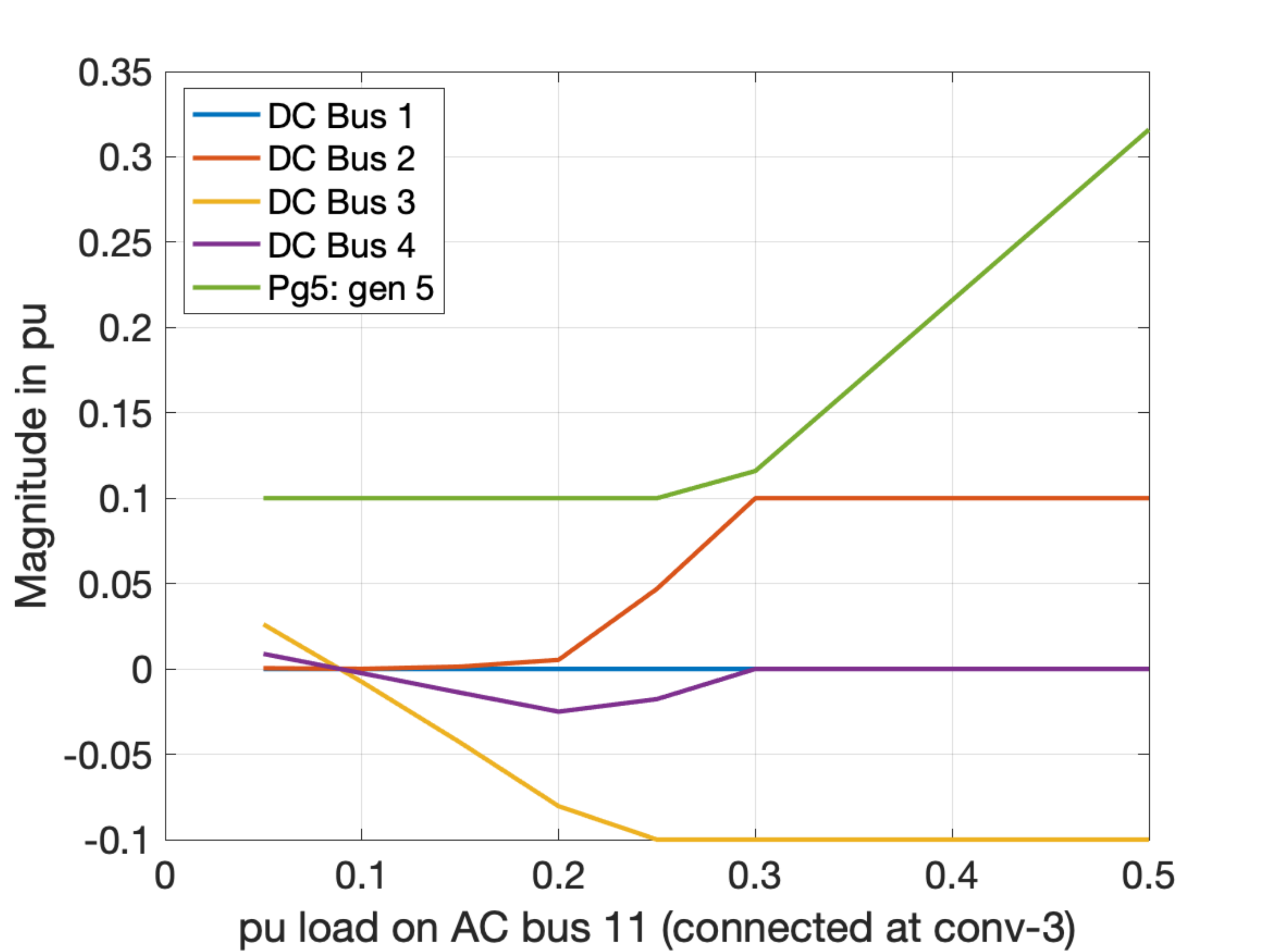}
        \caption{Impact of the higher metallic return resistance on the neutral terminal (DC bus) voltages and active power output of generator 5 in dependence of the unbalanced load at converter 3}
        \label{fig:load_variation}
    \end{figure}

The test case under study considers the values of the metallic return resistance equal to the value of pole (positive or negative) conductors. Since the metallic conductor is modeled separately, its resistance value {$\rdo$} can be chosen independently as per the system design and requirements. This value affects the system losses as well as the magnitudes of the neutral terminals of the DC buses. 
To show the impact of the metallic return conductor resistance more explicitly, it's value is chosen as $\rdo = 10 \cdot \rd$. Additionally, the load at AC bus 11, is varied from 0.05~$pu$ to 0.5~$p.u.$ with a step size of 0.05~$p.u.$, thus step-wise increasing unbalance in the system. For each step, the neutral terminal voltages of all DC buses and the output of generator 5 (connected at AC bus 11), are plotted in Fig. \ref{fig:load_variation}. It can be observed that until an unbalanced load of 0.25~$pu$, the output of the generator at AC bus 11 (Pg5) remains unchanged (at its lower limit of 0.1~$pu$). This means that the entire incremental load is supplied by the cheaper generators in the system through the DC grid. At this point, the neutral terminal voltage of DC bus 3 reaches its lower limit of -0.1~$pu$. Therefore, beyond this load point, Pg5 (the most expensive generator in the system) has to be dispatched more. At a load of 0.3~$pu$ at AC bus 11 (or DC bus 3), the neutral terminal voltage of DC bus 2 also reaches its limit (0.1~$pu$), which further limits the power transfer from the DC grid and yields a faster increase of the output of generator 5, indicated by the steeper slope of the curve. Thus, the flow of current over the metallic return conductor is constrained by both its resistance and neutral terminal voltages. Therefore, an explicit representation multi-conductor model is essential to capture these boundaries of the systems.


\section{Larger Test Cases and Computational Analysis}

\begin{table*}[h]
\centering
\caption{OPF results for various test cases with balanced bipolar configuration}
\label{tab: more_cases_balanced}
\begin{tabular}{l|ccccc|rr|rr}
\hline
\multirow{2}{*}{Test case} & \multicolumn{2}{c}{AC side} & \multicolumn{2}{c}{DC side} & \multicolumn{1}{l}{AC/DC} & \multicolumn{2}{c}{objective value} & \multicolumn{2}{c}{Solve time} \\
 & Bus & \multicolumn{1}{l}{Branches} & Bus & \multicolumn{1}{l}{Branches} & \multicolumn{1}{l}{Converters} & \multicolumn{1}{c}{$ACDC$} & \multicolumn{1}{c}{$MCDC$} & \multicolumn{1}{c}{$ACDC$} & \multicolumn{1}{c}{$MCDC$} \\ \hline
case5\_2grids\_MC & 11 & 14 & 4 & 3 & 3 & 861.2947 & 861.2947 & 0.0625 &0.1782 \\
case39\_mcdc & 39 & 46 & 10 & 12 & 10 & 41,995.5127 & 41,995.5127 & 0.4204 & 0.9961 \\
case67scopf\_mcdc & 67 & 102 & 9 & 11 & 9 & 86079.3816 & 86079.3816 & 0.2756 & 0.48069 \\
case3120sp\_acdc & 3,120 & 3693 & 5 & 5 & 5 & 2,142,635.0308 & 2142635.0308 & 19.15552 & 17.88725 \\ \hline
\end{tabular}
\end{table*}

\begin{table}[h]
\centering
\caption{OPF results for the large cases with unbalances on the DC side}
\label{Tab:more_cases_unbalance}
\begin{tabular}{lcp{0.4cm}rl}
\hline
\multirow{2}{*}{Test case} & \multicolumn{2}{c}{Objective value} & \multicolumn{1}{c}{} & \multicolumn{1}{c}{Outage}\\
 & \multicolumn{1}{c}{$ACDC$} & \multicolumn{1}{c}{$MCDC$} & \multicolumn{1}{c}{Solve time} & \multicolumn{1}{c}{on pole}\\ \hline
case5\_2grids\_MC & x & 886.2829 & 0.26028 & Conv-1 +Ve\\
case39\_mcdc & x & {41995.8283} & 1.40764 & Dcline-2 -Ve\\
case67\_mcdc & x & 86169.9679 & 0.89440 & Conv-1 -Ve\\
case3120sp\_acdc & x & 214288.1372 & 13.01446 & Conv-2 -Ve\\ \hline
\end{tabular}
\end{table}
 
The test case presented in the previous section is designed to explain the proposed multi-conductor modeling and demonstrate its value in capturing the system boundaries, i.e., constraints. However, the computational cost of the added detail in the proposed multi-conductor model and the scalability of the model are analyzed in this section considering that the modelling of the HVDC substations and the DC grid require a higher number of variables and constraints. The non-linear optimization problem is solved using the Ipopt solver in Julia/JuMP environment. All simulations are performed on a machine with a ``2.3 GHz 8-Core Intel Core i9" processor and ``32 GB 2667 MHz DDR4" RAM. 

Table \ref{tab: more_cases_balanced} presents the objective function values and the time taken to solve the OPF problem for test cases of different system sizes. Moreover, the DC side of all the test cases is chosen as a balanced bipolar system so that the accuracy and speed of the proposed multi-conductor model ($MCDC$) can be compared with the pre-existing single conductor model ($ACDC$, \cite{PMACDC.jl}). We can observe that the $MCDC$ converges to the same solution (objective function value) as the balanced single conductor model  up to the numerical accuracy of the solver. However, we observe that the solving time is higher than for the balanced model. Since the $MCDC$ model changes only the number of variables and constraints for the DC size, the changes in problem size and solve time are also related to the size of the DC grid. Here, the DC grid in the 67-bus test case is larger than the DC grid of the 3120-bus network. Therefore, the solution time of the 11-bus, 39-bus, and 67-bus increases to a larger extent for the unbalanced model, whereas for the 3120-bus system, it's almost the same ( or a bit less). Thus, as expected, the detailed modeling comes with an additional computational cost which obviously depends on the relative size of the DC system.  \\
Moreover, the value created by this additional cost is reflected more clearly in table \ref{Tab:more_cases_unbalance}, where the balanced model fails to capture the correct operating points for the AC/DC system with an unbalanced DC grid. From this table, it can be concluded that the proposed multi-conductor model converges to a feasible optimal solution for an unbalanced system, where the source of the unbalance is mentioned as a remark. As expected, the cost of operating the system in an unbalanced state (due to the corresponding component outage) is higher than that in a balanced case. 

\section{Conclusion}

A hybrid AC/DC optimal power flow model for the unbalanced operation of HVDC grids is introduced using a multi-conductor representation of the DC grid. The positive and negative poles of bipolar converter stations are modeled separately, and DC buses are modeled using three terminals: the positive, the negative, and the neutral terminal. Each conductor of a DC branch is modeled separately, including the metallic and ground return conductors. The proposed model is implemented as a software tool in the Julia language and validated against existing balanced AC/DC grid OPF models. 

It has been demonstrated that the unbalanced operation of HVDC grids cannot be correctly analyzed using balanced OPF models, as the assumption of equal power sharing among converter poles results in solutions that are infeasible when applied to the system. Further, the proposed multi-conductor representation can accurately capture such unbalances and provides converter and generator set points respecting the maximum allowable neutral terminal voltage, e.g., a maximum system unbalance. The numerical results show that in such unbalanced models, the explicit modeling of the converter current balance using an I-V formulation is essential due to the numerical problems caused by the near-zero values of neutral voltages in (P-V) formulations.

The proposed model and implementation provide a basis that will be extended in future work to incorporate further planning and operational aspects of hybrid AC/DC grids with unbalanced DC sides. 



\bibliographystyle{IEEEtran}
 \IEEEtriggeratref{31}
\bibliography{Referencelist_mcdc_opf}

\end{document}